\documentclass[12pt, a4paper]{iopart}
\usepackage{graphicx}
\usepackage{cite}
\usepackage{iopams}
\usepackage{dutchcal}

\newcommand{\avg}[1]{\langle{#1}\rangle}

\begin{document}

{\it \hfill Dedicated to the memory of Tobias Brandes}

\title{Interplay of coherent and dissipative dynamics in condensates of light}

\author{Milan Radonji\'{c}$^{1,2,3}$, Wassilij Kopylov$^4$, Antun Bala\v{z}$^1$ and Axel Pelster$^3$}
\address{$^1$Scientific Computing Laboratory, Center for the Study of Complex Systems, Institute of Physics Belgrade,
University of Belgrade, Serbia\\
$^2$Faculty of Physics, University of Vienna, Austria\\
$^3$Physics Department and Research Center OPTIMAS, Technische Universit\"at Kaiserslautern, Germany\\
$^4$Institute for Theoretical Physics, Technische Universit\"at Berlin, Germany}
\ead{milan.radonjic@ipb.ac.rs}

\begin{abstract}
Based on the Lindblad master equation approach we obtain a detailed microscopic model of photons in a dye-filled cavity, which features condensation of light. To this end we generalise a recent non-equilibrium approach of Kirton and Keeling such that the dye-mediated contribution to the photon-photon interaction in the light condensate is accessible due to an interplay of coherent and dissipative dynamics. We describe the steady-state properties of the system by analysing the resulting equations of motion of both photonic and matter degrees of freedom. In particular, we discuss the existence of two limiting cases for steady states: photon Bose-Einstein condensate and laser-like. In the former case, we determine the corresponding dimensionless photon-photon interaction strength by relying on realistic experimental data and find a good agreement with previous theoretical estimates. Furthermore, we investigate how the dimensionless interaction strength depends on the respective system parameters.
\end{abstract}


\section{Introduction}
\label{sec:intro}

Within the last decades open dissipative many-body quantum systems have emerged as a promising research direction for both basic research and applications. In particular, this is due to the development of exquisite technologies to coherently manipulate and control the internal and external degrees of freedom of atomic and photonic matter, as well as their interaction. A prominent example at the immediate interface of quantum optics and condensed matter physics is provided by the laser as a coherent light source which has contributed not only to our modern understanding of non-equilibrium phase transitions in general but even to many useful applications in our everyday life \cite{Haken,Scully,Siegman,Meschede}. 

Another more modern prominent object of research is the Bose-Einstein condensate (BEC) of light, which has so far been realised in a dye-filled micro-cavity at room temperature
in Bonn \cite{Klaers_BEC_of_photons}, in London \cite{Marelic_Photon_BEC_London}, and quite recently also in Utrecht \cite{NewOosten_1}. One of the key ingredients is the possibility of photons to acquire an effective mass by trapping them in a cavity in two dimensions -- without this, the photons would just disappear  according to the Planck law upon lowering the temperature. In the experiment, this is achieved via a curved-mirror cavity, which changes the dispersion relation of the photons from linear to quadratic. Along the resonator axis the frequency of the mode is quantised according to the resonance condition. Simultaneously, the curved mirrors create a harmonic trapping potential for the photons in the transverse direction. The next crucial element is given by dye molecules in the resonator, which are pumped incoherently. The multiple absorption and emission events between the cavity photons and the dye molecules lead to a thermalisation of the light \cite{Weitz-Thermalization}, so the resulting photon BEC emerges from an equilibrium phase transition \cite{FieldTheory,PhaseT-Critic_Properties_of_phi4_theory-Kleinert,PhaseT-renorm_group-zinn_justin}. Photon thermalisation was also shown to be possible in much simpler but periodically driven systems,  such as double quantum dots \cite{Gulans_thermalilzation_photons_double_qd_drive}, or a collection of harmonic modes \cite{Hafezi-chem_potential_for_light}, both coupled to some environment. {Recently, an elaborate theoretical proposal for a BEC of light in nanofabricated semiconductor micro-cavities has been put forward \cite{Stoof_Semiconductor_microcavities}.}

The usual atomic BEC as a thermal equilibrium phase transition occurs for temperatures below some critical value \cite{Pitaevskii-BEC,Pethick-Smith} {when the resulting ground state condensate acquires macroscopic occupation, while the populations of the higher energy levels obey the Bose-Einstein distribution even before the transition. Phase transition into a macroscopically occupied mode emerges as well in a controllable way in the laser case, but $-$ in contrast to the BEC $-$ this transition depends on the rate of the loss and the pumping channels, which makes it a non-equilibrium paradigm.} The two transitions, which {\it a priori} seem to be incompatible with each other due to their different nature, can thus be regarded as two sides of the same coin within a single non-equilibrium setup \cite{PhotonBEC-Thermalization_kinetics-Weitz}. Several studies concerning the similarities and differences of condensate and lasing states and their appearance in different systems exist
\cite{Bajoni-photon_lasing_similarities_with_condensate,Fischer_laser-and-bec-statistical_mechanic,Chioccetta-Laser_and_BEC-analog_and_diff,Leymann_Mode_Switching,Nyman_Small_phBEC,Nyman_Few_photon_BEC}. The investigation of systems and conditions under which a complex equilibrium state can be realized within a non-equilibrium setup has developed into an attractive topic, both in experiment and theory.

Apart from photonic systems, condensation effects of bosonic quasi-particles have also been observed in solid-state physics for magnons \cite{Magnons,Magnon-BEC-coherence,Magnon-BEC-pumping,Magnon-BEC-dipolar,Magnon-BEC-supercurrent} and exciton polaritons \cite{Kasprzak-BEC_of_polaritions,Malpuech-polarition_laser_thermo_vs_quantum_kinetic,Butov-polariton_laser,Kasprzak-Formation_excition_polarition_condensate,Byrnes_exciton-polarition-condensates,Ostrovskaya}. The latter quasi-particles  can be created in semiconductor micro-cavities  using strong coupling between photons and particle-hole excitations \cite{Kasprzak-BEC_of_polaritions}. A non-equilibrium BEC of polaritons has been observed in various experiments in polymers \cite{Plumhof-BEC_Polarition_Polymer_rommetemp,Stoeferle-exciton_bec_polymer}. Surprisingly, the transition there is not always restricted to a mode with the lowest momentum \cite{Richard-coherent_pt_of_polaritions}.

The first microscopic model of a photon condensate was developed by Kirton and Keeling \cite{Keeling_PRL-nonequilibrium_model_photon-cond,Keeling-Thermalization_photon_condensate}, which has recently been further extended by the same authors \cite{Keeling-Spatial_dynamics,Keeling-Polarization}. They considered a dye-filled cavity with multiple optical modes together with additional incoherent pump and loss channels and derived a Markovian quantum master equation of the Lindblad type \cite{Haake, Lindblad}.
Using an adiabatic elimination of the degrees of freedom of the dye molecules, Kirton and Keeling obtained a mean-field equation for the occupation of the cavity modes. The resulting steady state turned out to have different physical properties depending on the values of the respective system parameters. Provided that the relaxation time towards equilibrium is much shorter than the life time of the photons in the cavity, the steady state is given by a Bose-Einstein distribution, otherwise a laser-like state occurs having macroscopic occupation of a higher energy mode.
Inspired by such a behaviour, a minimal two-mode laser model with a Dicke-like interaction was investigated \cite{Brandes}. Different phases with up to four possible and up to two stable fixed points were found, some of which have an analogy to the laser-to-condensate-like transition. However, this analogy is only quite limited due to the absence of a temperature scale in the model.
{Quite recently, by considering the full spatial dynamics of light \cite{Keeling-Spatial_dynamics} a rich non-equilibrium phase diagram featuring Bose-Einstein condensation, multimode condensation and lasing has been demonstrated \cite{Hesten_Decondensation}.} On the other side, by using the Schwinger-Keldysh formalism a Langevin field equation describing the dynamics of photons in a dye-filled cavity was obtained \cite{Stoof-Keldysh} and later utilised to study phase fluctuations \cite{Stoof-Fluctuations} and phase diffusion \cite{Stoof-Diffusion} in such systems. Moreover, a quantum Langevin model for non-equilibrium condensation of photons in planar microcavity devices was developed in \cite{Chiocchetta-qm_langevin_noneq_photon_condens} and recently extended to address pseudo-thermalisation in driven-dissipative non-Markovian open quantum systems \cite{Carusotto-Pseudo-thermalization}. A theoretical description of a photon condensate based on three-dimensional Maxwell equations, which are mapped via a paraxial approximation to a two-dimensional Schr\"{o}dinger equation, was suggested as well 
\cite{PhotonBEC-theoretic_descrip_based_on_maxwell-Nyman}. {We also note that a unified theory for excited-state, fragmented and equilibrium-like Bose condensation in pumped photonic many-body systems has recently been introduced in \cite{Eckardt_Unified_theory}.}

The theoretical modelling of dissipative condensates usually strives for a reduced description in terms of a mean-field approximation in the form of a complex-valued Gross-Pitaevskii equation, which explicitly takes into account gains and losses. It describes the system around this phase transition even in non-equilibrium
\cite{Wouters-noneq-BEC-of_exciton,Lagoudakis-votices_in_excition_condensate,Bobrovska_Polariton_condensate-coherence}.
Within equilibrium, a real-valued Gross-Pitaevskii equation is a standard tool to describe condensation effects 
\cite{Pitaevskii-BEC,Pethick-Smith,Gross-Structure_quantized_vortex_boson_systems,Pitaevskii-Vortex_imperfect_Bose_gas,Diver-BEC_chaotic_dynamics_optical_cavity,Chiocchetta-non-eq_condensate}. At the present stage, the Gross-Pitaevskii-like equation for a photon condensate can only be obtained by including a non-linear self-interaction into the model on a phenomenological level
\cite{Klaers_BEC_of_photons,PhotonBEC-theoretic_descrip_based_on_maxwell-Nyman,Chiocchetta-qm_langevin_noneq_photon_condens}. A more detailed investigation shows that this non-linear self-interaction of photons is mediated via the change of the refractive index of the dye molecules due to the mutual presence of the optical Kerr and the thermo-optical effect \cite{ApplPhys}. Due to dimensional reasons the effective photon-photon interaction strength $g$ in two spatial dimensions corresponds to a dimensionless number $\tilde{g}= g m / \hbar^2$ \cite{Dalibard}, which turns out to be of the order of $10^{-9} \!-\! 10^{-8}$ for the Kerr 
and $10^{-4}$ for the thermo-optic effect, respectively \cite{ApplPhys}. Based on the observed momentum- and position-resolved spectra and images of the photoluminescence from thermalised and condensed dye-microcavity photons, the upper bound  $\tilde{g}\lesssim 10^{-3}$ was obtained \cite{NymanStrength}. In addition, a theoretical investigation of the influence of photon-photon interaction on the number fluctuations in a BEC of light \cite{StoofPRL-Interaction} successfully explained the measurements \cite{PhotonBEC-Observation_of_statistics-Weitz} and estimated the range $\tilde{g}\sim 10^{-8}-10^{-7}$. Surprisingly, even a much higher value for the interaction $\tilde{g} \sim 10^{-2}$ was recently measured \cite{Oosten}.

In this paper we generalise the microscopic model of the photon BEC by Kirton and Keeling \cite{Keeling_PRL-nonequilibrium_model_photon-cond,Keeling-Thermalization_photon_condensate} such that the dye-mediated contribution to the photon-photon interaction strength becomes microscopically accessible due to an interplay of coherent and dissipative dynamics. To this end, in section \ref{sec:model} we work out in detail the underlying model and discuss its improvements in comparison to \cite{Keeling_PRL-nonequilibrium_model_photon-cond,Keeling-Thermalization_photon_condensate}. Based on the corresponding
Lindblad master equation, we derive the resulting equations of motion of expectation values of the relevant system operators. In section \ref{sec:parameters} we determine the realistic model parameters in relation to current experiments. We then proceed in section \ref{sec:regimes} to analyse the steady-state properties of the system and identify the two limiting cases: a photon Bose-Einstein condensate and a laser-like regime. {The latter one is novel and accessible precisely due to the inclusion of the coherent dynamics.}
For the former case, in section \ref{sec:interaction} we determine the dye-mediated dimensionless photon-photon interaction strength from realistic experimental data and, in particular,
how it depends on the respective system parameters. Section \ref{sec:con} presents our concluding remarks.

\section{Model}
\label{sec:model}

Let us now introduce a physical system which encompasses both laser and photon BEC as the possible limiting cases. This is going to be done in close correspondence with the actual experimental setups of photon BEC experiments. We consider $N$ identical non-interacting two-level systems (TLS) inside an optical cavity. The transition between the two levels has the frequency $\Delta$ and it is nearly resonant with $M$ modes of the cavity. The dipole coupling between the TLS and the cavity modes has the strength $\mathcal{g}$ and it is assumed to be sufficiently weak so that the rotating wave approximation (RWA) holds. In the photon BEC experiments, the TLS were actually dye molecules dissolved in a solvent. The dye molecules have very broad rovibrational absorption and emission spectra, which can be modelled as an on-site phonon coupled to its own thermal bath \cite{Keeling_PRL-nonequilibrium_model_photon-cond,Keeling-Thermalization_photon_condensate}. {In addition, due to frequent collisions with the solvent particles the dye molecules experience rapid dephasing.} Hence, we take that each of the TLS is coupled to its own reservoir of $R\gg 1$ harmonic oscillators. This can be thought of as a compound reservoir consisting of a phonon and its bath. The reservoirs are supposed to be independent and of identical properties. {The collisional dephasing rate of each TLS is denoted by $\gamma_\phi$.} We also assume that the TLS are incoherently pumped to the excited state with the rate $\gamma_\uparrow$ and decay to the ground state with the rate $\gamma_\downarrow$ via spontaneous emission of photons outside of the cavity. The decay rate of all cavity modes is abbreviated by $\kappa$. A conceptually similar system has previously been treated by Kirton and Keeling \cite{Keeling_PRL-nonequilibrium_model_photon-cond,Keeling-Thermalization_photon_condensate} using a mixture of the master equation and the Schwinger-Keldysh formalisms{, but without accounting for the dephasing quantitatively}. Our approach, instead, is based entirely on the master equation formalism and we improve several aspects of their model. Later on we underline those specific points and our enhancements {that enable us to have access to a completely different regime of physical parameters.}

{In reality, the coupling between TLS and some cavity mode will also depend on the spatial mode function. A tractable model that incorporates the spatial dynamics was devised by Keeling and Kirton \cite{Keeling-Spatial_dynamics}. It has led to the successful understanding of the recent experiments \cite{Marelic_Photon_BEC_London,PhotonBEC-Thermalization_kinetics-Weitz}. However, the spatial dynamics introduces yet another level of complexity to the theoretical description. It could be implemented in our approach as well, but that would make the numerical calculations an order of magnitude more challenging. Thus, in the present work we make two additional simplifying assumptions: $(i)$ all TLS are at exactly the same position and $(ii)$ all cavity modes have the same intensity at the position of the TLS. This means that all TLS can be considered to evolve in an equivalent manner. Later on we will indicate how these assumptions may influence some of our results.}

\subsection{Master equation}

Due to the above mentioned assumptions we consider the system Hamiltonian ($\hbar=1$)
\numparts
\begin{eqnarray}
& H = \sum_{m=1}^M\omega_m a_m^\dag a_m^{} + \sum_{j=1}^N H_{\rm \uparrow\downarrow\:\!,R}^{(j)} + V^{}_{\rm \uparrow\downarrow\:\!,C},\label{Ham-a}\\
& H_{\rm \uparrow\downarrow\:\!,R}^{(j)} = \frac{\Delta}{2}\sigma_j^z +\sum_{r=1}^R\left[w_r b_{j,r}^\dag b_{j,r}^{}
+\lambda_r(b_{j,r}^\dag+b_{j,r}^{})\sigma_j^z\right],\label{Ham-b}\\
& V^{}_{\rm \uparrow\downarrow\:\!,C} = \mathcal{g}\sum_{m=1}^M\sum_{j=1}^N\left(a_m^\dag\sigma_j^- + a_m^{}\sigma_j^+\right),\label{Ham-c}
\end{eqnarray}
\endnumparts
where $\omega_m$ denote the cavity-mode frequencies and $a_m^{}$ ($a^\dag_m$) the bosonic annihilation (creation) operators of the cavity modes. The Hamiltonian $H_{\rm \uparrow\downarrow\:\!,R}^{(j)}$ describes the $j$-th TLS and its reservoir, with $\sigma_j^\pm$ and $\sigma_j^z$ being its Pauli spin operators. Bosonic annihilation (creation) operators and frequencies of the reservoir oscillators are $b_{j,r}^{}$ ($b_{j,r}^\dag$) and $w_r$, respectively, while $\lambda_r$ are the appropriate interaction strengths. Since the experimental spectra of the dye molecules are very broad, we are led to assume that the TLS-reservoir coupling is strong. In order to treat it non-perturbatively to all orders, we perform the polaron transformation $\tilde H=UHU^\dag$ with
\begin{equation}
U=\exp\left[\sum_{j=1}^N\sigma_j^z\sum_{r=1}^R \frac{\lambda_r}{w_r}(b_{j,r}^\dag-b_{j,r}^{})\right],
\end{equation}
and find
\numparts
\begin{eqnarray}
& \tilde H =\sum_{m=1}^M\omega_m a_m^\dag a_m^{}+\sum_{j=1}^N\frac{\Delta}{2}\sigma_j^z + \tilde V^{}_{\rm \uparrow\downarrow\:\!,C}
+ \sum_{j=1}^N H^{(j)}_{\rm R},\label{Ham-pol-a}\\
& \tilde V^{}_{\rm \uparrow\downarrow\:\!,C}=\mathcal{g}\sum_{m=1}^M\sum_{j=1}^N\left(a_m^\dag\sigma_j^-D_j^- + a_m^{}\sigma_j^+D_j^+\right),\quad
H^{(j)}_{\rm R} = \sum_{r=1}^R w_r b_{j,r}^\dag b_{j,r}^{},\label{Ham-pol-b}
\end{eqnarray}
\endnumparts
up to constant terms, where
\begin{eqnarray}
\label{Ds}
  D_j^\pm= {\displaystyle \otimes_{r=1}^R } \exp\left[\pm \frac{2\lambda_r}{w_r}\, (b_{j,r}^\dag-b_{j,r}^{})\right]
\end{eqnarray}
are the polaron displacement operators of the $j$-th TLS. In this way, $\tilde V^{}_{\rm \uparrow\downarrow\:\!,C}$ captures the coupling of the TLS and the cavity modes which is dressed by the reservoir oscillators.

In order to proceed further, we assume that the oscillators in the polaron frame represent a bath in a thermal state at temperature $T$
\begin{equation}
\rho^{}_\beta=Z_\beta^{-1}\otimes_{j=1}^N\exp\Big[-\beta H^{(j)}_{\rm R}\Big],
\end{equation}
where $Z_\beta$ stands for the canonical partition function and $\beta=1/(k_{\rm B} T)$ \cite{Schaller}.
We consider such initial conditions that the subsystem TLS-cavity is uncorrelated with the bath in the polaron frame, i.e., $\rho^{}_{\rm total}=\rho^{}_{\rm \uparrow\downarrow\:\!,C}\otimes\rho^{}_\beta$. Since the coupling strength $\mathcal{g}$ is supposed to be weak, the bath influence can be incorporated by means of a master equation, i.e., by treating $\tilde V^{}_{\rm \uparrow\downarrow\:\!,C}$ as a perturbation up to the second order \cite{JCP.94.4809,PRB.65.235311}. The first order contributes to the coherent unitary evolution through the thermal-averaged term
\begin{equation}
\langle\tilde V^{}_{\rm \uparrow\downarrow\:\!,C}\rangle_\beta = \mathcal{g}\sum_{m=1}^M\sum_{j=1}^N \left(a_m^\dag\sigma_j^-\langle D_j^-\rangle_\beta
+a_m^{}\sigma_j^+\langle D_j^+\rangle_\beta\right),
\end{equation}
where we introduced the notation $\langle X\rangle_\beta\equiv{\rm Tr}[X\rho^{}_\beta]$ for a bath expectation value. Using the result
\begin{equation}\label{eq:Davg}
\left\langle\exp\big[\alpha b^\dag-\alpha^* b\big]\right\rangle_\beta =
\exp\left[-\frac{|\alpha|^2}{2}\coth\frac{\beta\omega}{2}\right],
\end{equation}
for a harmonic oscillator of frequency $\omega$ in a thermal state, we find for the bath expectation value of the polaron displacement operators (\ref{Ds})
\begin{equation}\label{eq:Dpm}
\langle D_j^\pm\rangle_\beta = \exp\left[-2\sum_{r=1}^R\frac{\lambda_r^2}{w_r^2}
\coth\frac{\beta w_r}{2}\right]\,.
\end{equation}
Hence, one can naturally introduce a bath-dressed TLS-cavity coupling strength $\mathcal{g}_\beta=\mathcal{g}\langle D_j^\pm\rangle_\beta$. Obviously, due to (\ref{eq:Dpm}) we have $0<\mathcal{g}_\beta/\mathcal{g}<1$, so that the influence of the bath in the first order is to effectively reduce the TLS-cavity interaction
\begin{equation}\label{eq:V_A,C}
\langle\tilde V^{}_{\rm \uparrow\downarrow\:\!,C}\rangle_\beta = \mathcal{g}_\beta\sum_{m=1}^M\sum_{j=1}^N \left(a_m^\dag\sigma_j^- + a_m^{}\sigma_j^+\right).
\end{equation}
At this point we note that the previous first-order term was omitted by Kirton and Keeling \cite{Keeling_PRL-nonequilibrium_model_photon-cond,Keeling-Thermalization_photon_condensate}{, based on the implicit assumption that it is irrelevant due to the rapid collisional dephasing \cite{KK_Private_comm}}. As we will demonstrate below, its influence deep in the photon BEC regime turns out to be negligible, so this regime can be described satisfactorily even if it is not taken into account. However, in the opposite laser-like regime such a term does play a major role, {even in the presence of a fast sub-picosecond dephasing.} Anyhow, on formal grounds, it should be a part of the proper treatment.

We continue by applying the Born-Markov approximation as well as RWA, by tracing out the bath degrees of freedom and by taking into account the cavity losses along with the pumping and the decay of the TLS, similarly as \cite{Keeling_PRL-nonequilibrium_model_photon-cond,Keeling-Thermalization_photon_condensate}. {As already mentioned, we additionally account for the dephasing of the individual TLS.} Incoherent pumping can be formally described as coupling each TLS to a bath of inverted harmonic oscillators \cite{Quantum-Noise}. With this we find that the reduced density matrix $\rho^{}_{\rm \uparrow\downarrow\:\!,C}$ of the TLS-cavity subsystem obeys the following master equation
\begin{eqnarray}\label{eq:FullME}
\hspace*{-0.5cm}
\dot\rho^{}_{\rm \uparrow\downarrow\:\!,C}=&-\,i\left[\sum_{m=1}^M\delta_m a_m^\dag a_m^{}
+ \mathcal{g}_\beta \sum_{m=1}^M\sum_{j=1}^N \!\left(a_m^\dag\sigma_j^-\! + a_m^{}\sigma_j^+\right), \,\rho^{}_{\rm \uparrow\downarrow\:\!,C}\right]\nonumber\\
\hspace*{-1.5cm}
&-\left\{\sum_{m=1}^M\frac{\kappa}{2}{\cal L}[a_m]
+\sum_{j=1}^N\left(\frac{\gamma_\uparrow}{2}{\cal L}[\sigma_j^+] + \frac{\gamma_\downarrow}{2}{\cal L}[\sigma_j^-] +
{\frac{\gamma_\phi}{2}{\cal L}[\sigma_j^z]}\right)\right.\nonumber\\
\hspace*{-1.5cm}
&\left.+\sum_{m=1}^M\sum_{j=1}^N\left(\frac{\gamma^+_m}{2}{\cal L}[a_m^{}\sigma_j^+] + \frac{\gamma^-_m}{2}{\cal L}[a_m^\dag\sigma_j^-]\right) \!\right\}\rho^{}_{\rm \uparrow\downarrow\:\!,C},
\end{eqnarray}
where ${\cal L}[X]\rho = \{X^\dag X,\rho\}-2X\rho X^\dag$ and we have moved into the frame rotating with the frequency $\Delta$, so that $\delta_m=\omega_m-\Delta$ stands for the detuning of the cavity mode from the TLS transition. The thermal fluctuations of $\tilde V^{}_{\rm \uparrow\downarrow\:\!,C}$ give rise in the second order of perturbation theory to the incoherent transitions described by the dissipative Lindblad terms contained in the last double-sum of (\ref{eq:FullME}). The terms proportional to $\gamma^+_m$ correspond to the absorption of the cavity photons by the TLS, while those with the prefactor $\gamma^-_m$ represent the stimulated emission into the cavity modes. The previous approach should be satisfactory whenever $\tilde V^{}_{\rm \uparrow\downarrow\:\!,C}$ has small fluctuations around its thermal average and when the characteristic time scale, in which the bath modes undergo a displacement in order to adjust themselves to the instantaneous state of the TLS-cavity subsystem, is very short in comparison with the time scale of the subsystem relaxation. Note that the additional Lamb shifts due to the presence of the bath have been neglected as in \cite{Keeling_PRL-nonequilibrium_model_photon-cond,Keeling-Thermalization_photon_condensate}. Due to the dynamical influence of the bath, the corresponding rates $\gamma^\pm_m=\gamma(\pm\delta_m)$ turn out to be frequency dependent and are obtained along the lines of \cite{Keeling_PRL-nonequilibrium_model_photon-cond,Keeling-Thermalization_photon_condensate,PRL.107.093901} as
\begin{equation}\label{eq:gamma}
\gamma(\delta)=2\;\!\mathcal{g}^2\;\!{\rm Re}\int_0^\infty e^{-\frac{1}{2}(\gamma_\uparrow+\gamma_\downarrow)t}\;\!{\cal C}_\beta(t)\;\!e^{i\delta t}dt,
\end{equation}
with
\begin{equation}\label{eq:C}
{\cal C}_\beta(t)=\langle D_j^-(t)D_j^+\rangle_\beta-\langle D_j^-(t)\rangle_\beta\langle D_j^+\rangle_\beta
\end{equation}
being the retarded connected correlation function of the bath displacement operators. One can notice that the pumping and the decay of the TLS yield an exponentially decaying factor in (\ref{eq:gamma}), i.e., they introduce an additional level broadening \cite{PRL.107.093901}. The time evolution of $D_j^-(t)$ is generated by the free Hamiltonian $H^{(j)}_{\rm R}$, starting from $D_j^-(0)\equiv D_j^-$. Having in mind the result (\ref{eq:Davg}), one gets $\langle D_j^-(t)\rangle_\beta=\langle D_j^-\rangle_\beta$ and
\begin{equation}
  \label{DD}
\hspace*{-1cm}
\left\langle D_j^-(t)D_j^+\right\rangle_\beta = \exp\left\{-4\sum_{r=1}^R\frac{\lambda_r^2}{w_r^2}
\left[\left(1-\cos w_r t\right)\coth\frac{\beta w_r}{2}+i\sin w_r t\right]\right\}\, .
\end{equation}
Note that in the long-time limit the many oscillatory terms from the above sum simply add up to zero, such that, recalling (\ref{eq:Dpm}), one finds $\lim\limits_{t\to\infty}\langle D_j^-(t)D_j^+\rangle_\beta=\langle D_j^-\rangle_\beta\langle D_j^+\rangle_\beta$ and $\lim\limits_{t\to\infty}{\cal C}_\beta(t)=0$, i.e., the two displacement operators of very distant moments in time become uncorrelated.
In the Kirton-Keeling's approach \cite{Keeling_PRL-nonequilibrium_model_photon-cond,Keeling-Thermalization_photon_condensate}, the definition of the quantity (\ref{eq:gamma}) was actually without the last term of (\ref{eq:C}), i.e., ${\cal C}_\beta(t)$ had a finite long-time limit. On formal grounds, if both $\gamma_\uparrow$ and $\gamma_\downarrow$ are zero, that leads to a divergence of the absorption and emission rates of resonant light, i.e., $\gamma(\delta=0)\to\infty$. We trace this shortcoming back to the very absence of the first-order coherent term (\ref{eq:V_A,C}) from their treatment. Thus, the two improvements we have made to their approach come in pair.

The full master equation (\ref{eq:FullME}) is notoriously difficult to solve. However, its structure already reveals some general features of the system dynamics. Namely, one can clearly distinguish the coherent and the dissipative influence of the oscillator bath. The former one comes from the TLS-cavity coupling of the reduced strength $\mathcal{g}_\beta$ -- in a typical laser-like fashion, while the latter one is realised through the terms containing $\gamma_m^\pm$, which were shown to lead to thermalisation of light and emergence of photon BEC \cite{Keeling_PRL-nonequilibrium_model_photon-cond,Keeling-Thermalization_photon_condensate}. In the following we will demonstrate that precisely their interplay determines these two limiting stationary behaviours, i.e., a photon BEC or a laser.

\subsection{Equations-of-motion approach}

In order to be able to perform a quantitative analysis, we proceed to obtain the equations of motion for the mean values of the system observables $\avg{X}\equiv{\rm Tr}[X\rho^{}_{\rm \uparrow\downarrow\:\!,C}]$, e.g., the populations of the cavity modes, the population inversion of the TLS etc.\ from the master equation (\ref{eq:FullME}). Since this procedure yields an infinite hierarchy of coupled equations, we use the cumulant expansion method \cite{JPSJ.17.1100,PRA.82.033810,PRB.89.085308,Foerster_Computer-aided_cluster_expansion} to truncate the hierarchy at the second level, i.e., we will keep the cumulants up to the second order only. If one wants to calculate higher-order correlation functions, a higher level of truncation will be necessary. However, due to the presence of coherent terms in the master equation, the situation becomes considerably more involved than in \cite{Keeling_PRL-nonequilibrium_model_photon-cond,Keeling-Thermalization_photon_condensate}, even at this
second-order truncation level.

We note that the system possesses a $U(1)$ gauge symmetry: $a^{}_m\to a^{}_m e^{-i\phi}$, $a^\dag_m\to a^\dag_m e^{i\phi}$, $\sigma^-_j\to \sigma^-_j e^{-i\phi}$ and $\sigma^+_j\to \sigma^+_j e^{i\phi}$, $\phi\in\mathbb{R}$. In a single experimental run a coherent field with a particular spontaneously chosen phase $\phi$ can build up. However, since the density matrix describes an average over many such realisations, the following equalities hold $\avg{a_m^{}}=\avg{a_m^\dag}=\avg{\sigma^-_j}= \avg{\sigma^+_j}=0$ and similarly for all other gauge non-invariant operators \cite{PRA.82.033810}. In particular, it means that $\avg{a_m^{}\sigma^+_1}= \avg{a_m^{}\sigma^+_1}_{\rm c}$, $\avg{a^\dag_k a_m^{}}=\avg{a^\dag_k a_m^{}}_{\rm c}$ etc., since $\avg{XY}=\avg{XY}_{\rm c}+\avg{X}\avg{Y}$, where the index c denotes connected correlation functions. For instance, one has
\begin{eqnarray}
\hspace*{-1.5cm}
\avg{a^\dag_k a_m^{} \sigma^z_1} = &\avg{a^\dag_k a_m^{}\sigma^z_1}_c + \avg{a^\dag_k a_m^{}}_c \avg{\sigma^z_1} + \avg{a^\dag_k \sigma^z_1}_c \avg{a_m^{}}+\avg{a_m^{} \sigma^z_1}_c \avg{a^\dag_k}\nonumber\\
 \hspace*{-1.5cm}
 &+ \avg{a^\dag_k}\avg{a_m^{}}\avg{\sigma^z_1} \approx \avg{a^\dag_k a_m^{}} \avg{\sigma^z_1}.
\end{eqnarray}
{Due to the two simplifying assumptions mentioned in the beginning,} we assume that all the TLS are mutually equivalent such that, e.g., $\avg{\sigma^z_i}=\avg{\sigma^z_1}$ and $\avg{\sigma^+_i\sigma^-_j}=\avg{\sigma^+_1\sigma^-_2}$ for all $i,j=1,\ldots,N$ and $i\neq j$. The resulting equations of motion
for the cavity mode occupations $\avg{n_m^{}}\equiv\avg{a^\dag_m a_m^{}}$ read
\begin{eqnarray}
\hspace*{-1.5cm}
\frac{d}{dt}\avg{n_m^{}}=&-\kappa\avg{n_m^{}}+iN \mathcal{g}_\beta\left(\avg{a_m^{}\sigma^+_1}-\avg{a^\dag_m\sigma^-_1}\right)
-\frac{N}{2}\gamma^+_m\avg{n_m^{}}(1-\avg{\sigma^z_1})
 \nonumber\\\hspace*{-1.5cm}
&+\frac{N}{2}\gamma^-_m(\avg{n_m^{}}+1)(1+\avg{\sigma^z_1})\,,\label{eq:MainSys-1}
\end{eqnarray}
whereas for the TLS population inversion $\avg{\sigma^z_1}$ we obtain
\begin{eqnarray}
\hspace*{-1.5cm}
\frac{d}{dt}\avg{\sigma^z_1}=&\gamma_\uparrow\left(1-\avg{\sigma^z_1}\right)
-\gamma_\downarrow\left(1+\avg{\sigma^z_1}\right)
+2i\mathcal{g}_\beta\sum_{m=1}^M\left(\avg{a^\dag_m\sigma^-_1}
-\avg{a_m^{}\sigma^+_1}\right)
\nonumber\\\hspace*{-2.0cm}
&+\sum_{m=1}^M\left[\gamma^+_m\avg{n_m^{}}\left(1-\avg{\sigma^z_1}\right)-\gamma^-_m\left(\avg{n_m^{}}+1\right)\left(1+\avg{\sigma^z_1}\right)\right].\label{eq:MainSys-2}
\end{eqnarray}
These equations are exactly like those of Kirton and Keeling \cite{Keeling_PRL-nonequilibrium_model_photon-cond,Keeling-Thermalization_photon_condensate}, apart from the coherent terms proportional to $\mathcal{g}_\beta$ which introduce the additional coupling to the mixed-type terms $\avg{a_m^{}\sigma^+_1}=\avg{a^\dag_m\sigma^-_1}^*$. They measure the correlation between TLS and cavity photons and evolve according to
\begin{eqnarray}
\hspace*{-2cm}
\frac{d}{dt}\avg{a_m^{}\sigma^+_1}=&\left(-i\delta_m-\frac{\gamma_\uparrow+\gamma_\downarrow+4\gamma_\phi+\kappa}{2}\right)\avg{a_m^{}\sigma^+_1}\nonumber\\
\hspace*{-2cm}
&-i\mathcal{g}_\beta\left[\frac{1+\avg{\sigma^z_1}}{2}+\avg{\sigma^z_1}\sum_{k=1}^M\avg{a^\dag_k a_m^{}}+(N-1)\avg{\sigma^+_1\sigma^-_2}\right]\nonumber\\
\hspace*{-2cm}
&-\frac{\gamma^+_m}{4}(N-1)\avg{a_m^{}\sigma^+_1}\left(1-\avg{\sigma^z_1}\right)
+\frac{\gamma^-_m}{4}(N-1)\avg{a_m^{}\sigma^+_1}\left(1+\avg{\sigma^z_1}\right)\nonumber\\
\hspace*{-2cm}
&-\sum_{k=1}^M\left\{\frac{\gamma^+_k}{2}\left[\avg{a_m^{}\sigma^+_1}\avg{n_k^{}}+\avg{a_k^{}\sigma^+_1}{\avg{a_m^{} a^\dag_k}}\right]\right.\nonumber\\
\hspace*{-2cm}
&\left.+\frac{\gamma^-_k}{2}\left[\avg{a_m^{}\sigma^+_1}\left(\avg{n_k^{}}+1\right)+\avg{a_k^{}\sigma^+_1}\avg{a^\dag_k a_m^{}}\right]\right\},\label{eq:MainSys-3}
\end{eqnarray}
where now the additional quantities $\avg{a_k^\dag a_m^{}}$ and $\avg{\sigma^+_1\sigma^-_2}$ appear. The former represent the correlations between different cavity modes, whose evolution is governed by
\begin{eqnarray}
\hspace*{-2.4cm}
\frac{d}{dt}\avg{a_k^\dag a_m^{}}=\left[-\kappa+i(\delta_k-\delta_m)\right]\avg{a^\dag_k a_m^{}}
+iN \mathcal{g}_\beta\left(\avg{a_m^{}\sigma^+_1}-\avg{a^\dag_k\sigma^-_1}\right)\nonumber\\
\hspace*{-0.2cm}
+\frac{N}{4}\left(\gamma^-_k+\gamma^-_m\right)\avg{a_m^{}a^\dag_k}\left(1+\avg{\sigma^z_1}\right)-\frac{N}{4}\left(\gamma^+_k+\gamma^+_m\right)\avg{a^\dag_k a_m^{}}\left(1-\avg{\sigma^z_1}\right),\label{eq:MainSys-4}
\end{eqnarray}
and the latter the correlations between dipoles of different TLS following from
\begin{eqnarray}
\hspace*{-2cm}
\frac{d}{dt}\avg{\sigma^+_1\sigma^-_2}=&-(\gamma_\uparrow+\gamma_\downarrow+{4\gamma_\phi})\avg{\sigma^+_1\sigma^-_2}
+i\mathcal{g}_\beta\avg{\sigma^z_1}\sum_{m=1}^M\left(\avg{a_m^{}\sigma^+_1}-\avg{a^\dag_m\sigma^-_1}\right)\nonumber\\
\hspace*{-2cm}
&-\sum_{m=1}^M\bigg\{\gamma^+_m\left[\avg{\sigma^+_1\sigma^-_2}\avg{n_m^{}}+\avg{a^\dag_m\sigma^-_1}\avg{a_m^{}\sigma^+_1}\right]\nonumber\\
\hspace*{-2cm}
&+\gamma^-_m\left[\avg{\sigma^+_1\sigma^-_2}\left(\avg{n_m^{}}+1\right)+\avg{a_m^{}\sigma^+_1}\avg{a^\dag_m\sigma^-_1}\right]\bigg\}\,.\label{eq:MainSys-5}
\end{eqnarray}
It is important to notice that the quantities $\avg{a_m^{}\sigma^+_1}$, $\avg{a_k^\dag a_m^{}}$ and $\avg{\sigma^+_1\sigma^-_2}$ can reach non-zero stationary values precisely due to the coherent part of the evolution, which we have introduced in addition to \cite{Keeling_PRL-nonequilibrium_model_photon-cond,Keeling-Thermalization_photon_condensate}.

At this point, one additional specialisation is in order. Namely, based on the photon BEC experiments, we consider the photon modes as being transverse, arising from a two-dimensional effective harmonic potential \cite{Klaers_BEC_of_photons}. We thus consider regularly spaced cavity levels $\omega_\ell=\omega_1+ (\ell-1)\Omega$ with $\ell=1,\ldots,L$, such that the energy level $\omega_\ell$ has the degeneracy $d_\ell^{}=2\ell$, where the factor 2 comes from the two independent polarisations of light. The lowest frequency $\omega_1$ represents the cavity cutoff. Since degenerate cavity modes evolve in the same manner, we have
\begin{equation}
\sum_{m=1}^M f(a^{}_m,a^\dag_m,\ldots)=\sum_{\ell=1}^L d_\ell^{}\, f(a^{}_\ell,a^\dag_\ell,\ldots),
\end{equation}
for any arbitrary function $f$, where from each level $\omega_\ell$ we have chosen a representative mode described by $a^{}_\ell$ and $a^\dag_\ell$.

\subsection{Bath model}

In the following we analyse the stationary solutions of the equations
(\ref{eq:MainSys-1})--(\ref{eq:MainSys-5}) in the two regimes: a photon BEC and a laser-like regime. To this end, we specialise the model by choosing the bath spectral density, defined by $J(w)=\sum_{r=1}^R\lambda_r^2\,\delta(w-w_r)$, to be {super-ohmic with an exponential cut-off \cite{Weiss_book,Lee_Polaron_frames}}
\begin{equation}\label{eq:J(w)}
J(w)=\frac{\eta}{w_c^2}w^3\exp\left(-w/w_c\right),
\end{equation}
where $\eta$ represents a dimensionless parameter measuring the coupling strength of the TLS to the bath and $w_c$ is the cutoff frequency of the bath. This allows us to obtain for (\ref{eq:Dpm}) and
(\ref{DD}) the closed-form expressions
\numparts
\begin{eqnarray}
\hspace*{-2.55cm}
\left\langle D_j^\pm\right\rangle_\beta = \exp\left\{2\eta\,\left[1-\frac{2\psi'\big(\frac{1}{\beta w_c}\big)}{(\beta w_c)^2}\right]\right\},\label{eq:Dm-Dp-eta-a}\\
\hspace*{-2.55cm}
\left\langle D_j^-(t) D_j^+\right\rangle_\beta = \exp\left\{4\eta\,\left[1-\frac{1}{(1-i w_c t)^2}+
\frac{\psi'\big(\frac{1-i w_c t}{\beta w_c}\big)+\psi'\big(\frac{1+i w_c t}{\beta w_c}\big)-2\psi'\big(\frac{1}{\beta w_c}\big)}{(\beta w_c)^2}
\right]\right\},\label{eq:Dm-Dp-eta-b}
\end{eqnarray}
\endnumparts
where $\psi(z)=\Gamma'(z)/\Gamma(z)$ denotes a logarithmic derivative of the gamma function. In relation to the experiments, both parameters $\eta$ and $w_c$ characterise the impact of the used solvent on the spectral properties of the dye molecules. Below we discuss in detail that they change the absorption and the emission spectra $\gamma(\pm\delta)$ drastically. Furthermore, they turn out to have an impact upon the resulting
bath-dressed coupling strength $\mathcal{g}_\beta$.

\section{Determination of realistic model parameters}
\label{sec:parameters}

In order to apply our theory to the current experimental setups, we have to fix the model parameters in the experimentally accessible regimes. One of the key ingredients for the thermalisation of photons are the spectral properties of the dye molecules \cite{PhotonBEC-Thermalization_kinetics-Weitz,Carusotto-Pseudo-thermalization}. Whereas the Einstein rate coefficient for absorption $B_{12}(\omega)$ is usually measured, the stimulated emission rate $B_{21}(\omega)$ is determined via the Kennard-Stepanov relation \cite{Kennard1,Kennard2,Stepanov}
\begin{equation}
\label{eq:Kennard-Stepanov}
\frac{B_{21}(\omega)}{B_{12}(\omega)} \sim \exp\left(-\,\frac{\hbar \omega}{k_{\rm B} T} \right)  \, ,
\end{equation}
where the spectral (rovibrational) temperature $T$ of the dye molecules is $T = 300~{\rm K}$ \cite{PhotonBEC-Thermalization_kinetics-Weitz}. Comparing the rate equation from the Supplemental Material of \cite{PhotonBEC-Thermalization_kinetics-Weitz} with our equation~(\ref{eq:MainSys-1}), we can interpret the rates $\gamma^+_m=\gamma(\delta_m)$ and $\gamma^-_m=\gamma(-\delta_m)$ as the absorption and the emission rates $B_{12}(\omega_m)$ and $B_{21}(\omega_m)$, respectively. Therefore, we fit the  expression $\gamma(\omega-\Delta)$ from (\ref{eq:gamma}), using equations (\ref{eq:Dm-Dp-eta-a}) and (\ref{eq:Dm-Dp-eta-b}) as well, to the experimentally measured absorption spectrum $B_{12}(\omega)$ of the used Rhodamine 6G dye dissolved in ethylene glycol \cite{Private-Julian_Schmitt}, by taking into account the absolute value $B_{12}(\omega = 3300~{\rm THz}) = 1.3~{\rm kHz}$ \cite{PhotonBEC-Thermalization_kinetics-Weitz}. The fitting allows then to fix the parameters of our bath model, namely the dimensionless coupling strength $\eta$ of the TLS and the bath, the cutoff frequency of the bath $w_c$, as well as the TLS-cavity coupling strength $\mathcal{g}$ and the TLS transition frequency $\Delta$, which physically corresponds to the zero-phonon-line frequency of the dye. The obtained values are listed in table~\ref{tab:exp_parameters}. Thus, the fitted value for $\eta = 0.6$ leads to a rather small bath-dressed TLS-cavity coupling strength $\mathcal{g}_\beta = 8.3\times 10^{-3}\,\mathcal{g}$, as expected, since this corresponds to the BEC regime.

\begin{table}[!b]
\caption{\label{tab:exp_parameters}
Parameters of the model adjusted to current experimental setup of photon BEC \cite{Klaers_BEC_of_photons}. The first four parameters are obtained from the fit to the absorption spectrum of the dye. The remaining parameters are taken from \cite{PhotonBEC-Thermalization_kinetics-Weitz,PhotonBEC-Observation_of_statistics-Weitz,Private-Julian_Schmitt}.}
\begin{indented}\setlength{\leftskip}{-6mm}
\item[]
\begin{tabular}{@{}lllllllll@{}}
\br
$w_c$ & $\eta$ & $\mathcal{g}$ & $\Delta$ & $\kappa$ & $\delta_1$ & $\delta_L$ & $T$ & $N$ \\
\mr
20.5~THz & 0.6 & 2.46~GHz & 3487~THz & 3.5~GHz & -260~THz & -120~THz & 300~K & $10^9$\\
\br
\end{tabular}
\end{indented}
\end{table}

\begin{figure}[!t]
\centering
\includegraphics[width=0.55\linewidth]{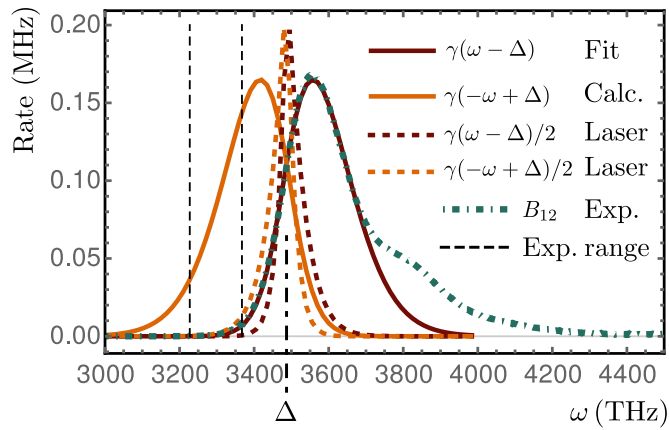}
\caption{Absorption rate $\gamma(\omega-\Delta)$ (brown/dark) fitted to measured data from \cite{PhotonBEC-Thermalization_kinetics-Weitz,Private-Julian_Schmitt} (dash-dotted). Emission rate $\gamma(-\omega+\Delta)$ (orange/light) for same parameters. Absorption and emission rates are also shown for lasing regime with $\eta = 0.1$ (dashed).}
\label{fig:rates_fit}
\end{figure}

Figure~\ref{fig:rates_fit} shows the resulting fit for $\gamma(\omega-\Delta)$ and the experimentally provided data. Note, that we do not fit the absorption curve for $\omega > 3700~{\rm  THz}$, since most of the relevant cavity modes are not influenced by the departure from the actual spectrum, as is indicated by the vertical dashed lines. If one wants to also fit this higher-frequency region, the inclusion of another bath would be necessary since the experimental spectrum displays the presence of another peak. A corresponding physical motivation in terms of another dye-molecule active phonon coupled to a thermal environment was discussed in \cite{Keeling-Thermalization_photon_condensate}. With the fitted values our theory provides the emission rate curve $\gamma(-\omega+\Delta)$ as well, which is shown in the same figure. The curves $\gamma(\omega-\Delta)$ and $\gamma(-\omega+\Delta)$ cross at the frequency $\Delta$. We checked that the Kennard-Stepanov relation (\ref{eq:Kennard-Stepanov}) is valid in the BEC regime for the relevant range of the cavity mode frequencies. For comparison, in the laser regime where $\eta$ is small, the absorption and emission curves become squeezed towards the zero-phonon-line frequency $\Delta$, see the dotted curves in figure~\ref{fig:rates_fit}. However, the Kennard-Stepanov relation (\ref{eq:Kennard-Stepanov}) in this regime is no longer granted for frequencies highly detuned from $\Delta$. The corresponding experimental number of dye molecules is taken to be $N=10^9$, based on the extensive discussion in \cite{PhotonBEC-Observation_of_statistics-Weitz}. The loss rate $\gamma_\downarrow$ can be fixed to $0.25~{\rm GHz}$ \cite{Private-Julian_Schmitt}. Furthermore, the pumping of the TLS $\gamma_\uparrow$  is considered as a control parameter which is tuned to cross the boundary of the phase transition. {Knowing that dye molecules experience at least $10^{12}$ collisions per second with solvent molecules \cite{Schaefer_Dye_Lasers}, we will consider here $\gamma_\phi=0.1~{\rm THz}$ as a referent value. Since there is an uncertainty about the exact order of magnitude of this rate, we will later analyse how its variation in a broad range of values influences our results.}

In the experiment the cavity cutoff wavelength, corresponding to the mode of frequency $\omega_1$, can be tuned from 570~nm to 610~nm. The frequency separation of the cavity modes is $0.26\,{\rm THz}$ \cite{Klaers_BEC_of_photons}. For the simulation we choose $\omega_1$ to correspond to $585~{\rm nm}$, thus we get for the highest detuning $\delta_1 = \omega_1 - \Delta = -260\,{\rm THz}$. Our simulations cover the same spectral range as in \cite{Klaers_BEC_of_photons}, but due to computational complexity we choose a higher value $\Omega= 1.4\,{\rm THz}$, if not stated otherwise. The photon loss rate $\kappa$ is frequency-dependent \cite{PhotonBEC-Thermalization_kinetics-Weitz}, so we take a mean value $\kappa = 3.5~{\rm GHz}$. For the sake of clarity the used spectral range is marked in figure~\ref{fig:rates_fit} by vertical dashed lines.

\section{Two regimes: photon BEC and laser-like state}
\label{sec:regimes}

Having discussed the realistic values of the model parameters, in this section we take: $N=10^9$, $w_c=20.5\,{\rm THz}$, $T=300\,{\rm K}$, $\mathcal{g}=2.3\,{\rm GHz}$, $\gamma_\downarrow=0.25\,{\rm GHz}$, $\gamma_\uparrow=0.1\,{\rm GHz}$, $\kappa=3.5\,{\rm GHz}$ and $\delta_1=-260\,{\rm THz}$. {The results will be presented for two values of the dephasing rate, $\gamma_\phi=0.1~{\rm THz}$ estimated from the literature \cite{Schaefer_Dye_Lasers} and a much larger one, $\gamma_\phi=10~{\rm THz}$, for the sake of comparison.} Here we take $\Omega=10\,{\rm THz}$ and consider $L=20$ energy levels of the cavity. In the following we distinguish two limiting regimes:
\begin{itemize}
\item[$(i)$] $\eta\gtrsim 1$. In this case one has $\mathcal{g}_\beta/\mathcal{g}=\langle D_j^\pm \rangle_\beta\ll 1$, meaning that the coherent contribution of the bath is highly suppressed and the evolution is dominated by the dissipative influence which leads to a thermalisation of light and an emergence of photon BEC;
\item[$(ii)$] $\eta\ll 1$. Here one finds $\mathcal{g}_\beta/\mathcal{g}\approx 1$, so that the bath has a pronounced coherent influence. In addition, the rates $\gamma_m^\pm$ of the highly detuned cavity modes acquire orders of magnitude smaller values
in comparison with the previous regime, so that the dissipative influence becomes overwhelmed, but is still relevant. Hence, we expect that the non-equilibrium stationary state is then highly coherent and laser-like.
\end{itemize}
\begin{figure}[!t]
\centering
\includegraphics[width=\linewidth]{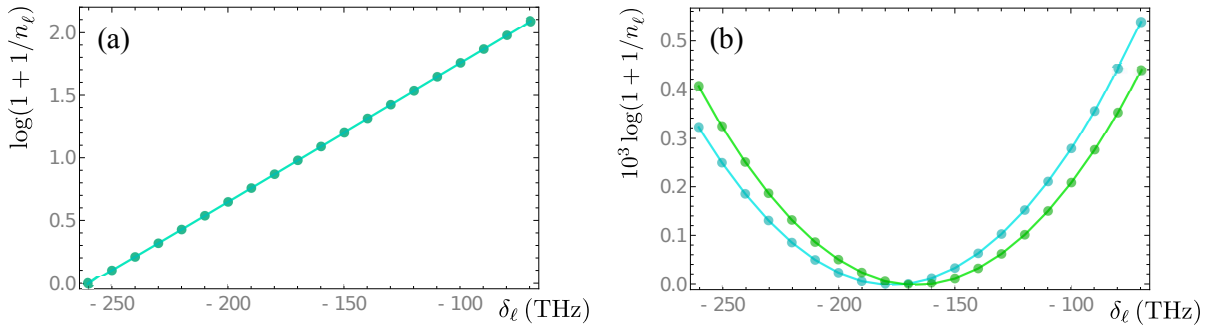}
\caption{Stationary occupations $n_\ell^{}$ of cavity modes $\ell=1,\ldots,20$ when (a) $\eta=1.5$ and (b) $\eta=0.05$. {Cyan colour corresponds to $\gamma_\phi=0.1\,{\rm THz}$, while green is used for $\gamma_\phi=10\,{\rm THz}$.} For other used parameters see main text. {Note that in (a) the two cases are visually indistinguishable.}}\label{fig2}
\end{figure}
For the regime $(i)$ we choose $\eta=1.5$, yielding $\mathcal{g}_\beta/\mathcal{g}=6.2\times 10^{-6}$. The resulting steady state solution yields the distribution of occupations of the cavity modes $n_\ell^{}$ for $\ell=1,\ldots,20$ as shown in figure~\ref{fig2}(a). {It is noticeable that the results (almost) do not depend on the value of dephasing rate. This is predictable since the coherent evolution is anyhow largely suppressed in this regime.} The lowest energy level is macroscopically occupied with about $2.84\times 10^7$ photons. The straight line corresponds to a Bose-Einstein distribution $\log(1+1/{n_\ell^{}})=\beta(\delta_\ell-\mu)$, where $\mu$ denotes the chemical potential. Such a state was already analysed in detail in \cite{Keeling_PRL-nonequilibrium_model_photon-cond,Keeling-Thermalization_photon_condensate} and it was shown to correspond to a photon BEC. Our approach enables us to additionally characterise the stationary states by their photonic correlations. Namely, we have access to the quantities
\begin{equation}
c_{\ell,\ell'}^{}=\frac{|\avg{a^\dag_\ell a_{\ell'}^{}}|}{\sqrt{\vphantom{1}{n_\ell^{}}\;\!{n_{\ell'}^{}}}},\quad 1\le \ell<\ell'\le L\, ,
\end{equation}
which provide a measure of correlations between representative cavity modes related to different energy levels. Their values belong to the interval $[0,1]$, where values close to 1 (0) correspond to a high (low) degree of correlation. In case  $(i)$ we find $c_{\ell,\ell'}^{}<4\times 10^{-7}$, i.e., the photon BEC state has almost no correlation between the modes of different frequencies. This is expected since the correlations build up through the coherent evolution which is highly ineffective in this regime.

In the opposite regime $(ii)$, we take $\eta=0.05$, which gives $\mathcal{g}_\beta/\mathcal{g}=0.67$. The distribution of stationary populations of the representative cavity modes is presented in figure~\ref{fig2}(b) {for two values of $\gamma_\phi$, namely $0.1\,{\rm THz}$ and $10\,{\rm THz}$.} In the former case, the cavity level $\ell=9$ acquires macroscopic occupation of almost $1.90\times 10^7$ photons, but the distribution is quite distinct from the Bose-Einstein one. In this case we find $c_{\ell,\ell'}^{} >0.996$, which demonstrates that the stationary state contains a quite high degree of correlations among the cavity modes of different energies. This is expected since the coherent influence of the bath is very pronounced. The stationary state is laser-like with some dissipative bath influence. {For the larger value of the dephasing rate, the level $\ell=10$ becomes macroscopically occupied with $1.76\times 10^7$ photons, while $c_{\ell,\ell'}^{}>0.986$. Interestingly, in the considered parameter regime the results for $\gamma_\phi=0$ would be almost indistinguishable from those for $\gamma_\phi=0.1\,{\rm THz}$. This means that there is a certain dephasing threshold below which the coherent system dynamics is robust to the dephasing. Moreover, if one considered the full spatial structure of the cavity modes as in \cite{Keeling-Spatial_dynamics}, the aforementioned correlations would decrease due to only partial overlap among different modes. However, this would not alter the present conclusions.} We note that similar states supporting macroscopic occupations of optical modes of higher energies were also observed in \cite{Keeling_PRL-nonequilibrium_model_photon-cond,Keeling-Thermalization_photon_condensate} and, indeed, we can also reproduce such behaviour in the regime $(i)$. However, the steady-state we have just presented features near-unity correlations of light, which represents a crucial difference and indicates that it is of an entirely different nature. Moreover, for different values of the cavity decay rate even the lowest level can acquire macroscopic occupation, while the populations of other cavity levels strongly depart from the Bose-Einstein distribution. Hence, the stationary states in the two regimes $(i)$ and $(ii)$ differ completely regarding the correlations and the distribution of photons among the cavity levels, as a consequence of different influences of the bath.

\section{Properties of photon BEC}\label{sec:interaction}

In the following, we focus on interesting properties of the photon BEC. At first, we determine from our model the equation of state. This allows then to extract the dimensionless effective photon-photon interaction strength in the photon BEC regime and study its dependence on various model parameters, which could be tuned experimentally. We will adopt the terminology in accordance with the experiments and, for instance, instead of TLS refer to dye molecules.

\subsection{Equation of state}

In this section we apply our theory to experimentally realistic values and determine at first the steady state of the equations of motion (\ref{eq:MainSys-1})--(\ref{eq:MainSys-5}) for different pumping rates $\gamma_\uparrow$ and evaluate the dependence of the chemical potential $\mu$ on the total photon number $n_{\rm tot}$, i.e., we obtain the equation of state $\mu(n_{\rm tot})$. In the following we present and discuss this procedure in detail by using the specific parameters from table~\ref{tab:exp_parameters}, if not stated otherwise. 

\begin{figure}[!t]
\centering
\includegraphics[width=\linewidth]{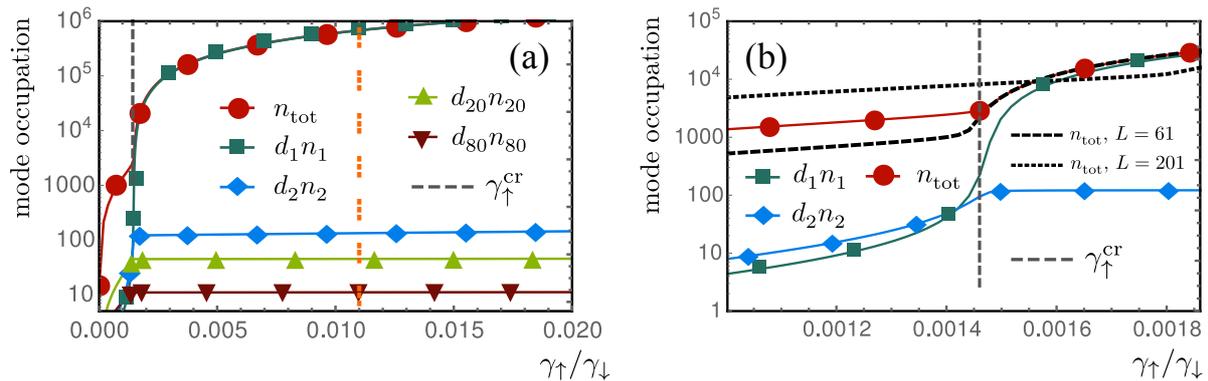}
\caption{(a) Occupation of 1st, 2nd and some higher cavity levels as well as total occupation of all modes as function of pumping $\gamma_\uparrow$ reveals condensation for $\gamma_\uparrow > \gamma_\uparrow^{\rm cr}$ (vertical dashed line) for $L=101$. Further increase of $\gamma_\uparrow$ mainly populates the lowest level. Orange triple-dashed vertical line marks the value $\gamma_\uparrow = 0.011~\gamma_\downarrow$.
(b) Zoom around $\gamma_\uparrow^{\rm cr}$ shows onset of condensation in the lowest energy level (green squares). Onset of condensation in total number of photons (red circles) becomes smoothened for higher number of levels (red circles vs.~dashed and dotted lines). Transition is also visible in occupation of higher levels (blue rhombi). Parameters are listed in table~\ref{tab:exp_parameters}.}
\label{fig:occ-mu-plot}
\end{figure}

In figure~\ref{fig:occ-mu-plot}(a) we show the occupation of different energy levels of the cavity in the BEC regime for the increasing pump rate $\gamma_\uparrow$ with all other parameters being fixed. The onset of the condensation starts at the critical value $\gamma_\uparrow^{\rm cr}$. A zoom around $\gamma_\uparrow^{\rm cr}$ is shown in figure~\ref{fig:occ-mu-plot}(b). Clearly, the occupation of the lowest level $d_1 n_1$ shows a sudden increase at the critical point and becomes macroscopic afterwards. Further increase of $\gamma_\uparrow$ mainly populates the lowest level, while the population of higher energy levels does not change significantly after the transition. At the critical value of the pumping parameter $\gamma_\uparrow^{\rm cr}$ the total number of photons $n_{\rm tot} = \sum_{\ell=1}^{L} d_\ell^{} n_\ell^{}$ amounts to $n_{\rm tot}(\gamma_\uparrow^{\rm cr}) \approx 2800$, which is quite close to the expected critical photon number for the condensation onset in the case of non-interacting interacting bosons in two dimensions \cite{Pethick-Smith,Pitaevskii-BEC} and at $T=300~{\rm K}$
\begin{equation}
\label{eq:phot_crit_cond}
n^{\rm cr} = \frac{\pi^2}{3} \, \left(\frac{k_{\rm B} T}{\hbar \Omega }\right)^2 = 2640,
\end{equation}
where the existence of two independent polarisations of light has been taken into account. Note that the rise of the total photon number $n_{\rm tot}$ around the transition point becomes smoothened when the number of cavity levels $L$ in the considered frequency range is increased, as is seen in figure~\ref{fig:occ-mu-plot}(b). This can be explained with the significant contribution of degeneracies $d_\ell^{}$ of the energy levels with high $\ell$ to $n_{\rm tot}$.

\begin{figure}[!t]
\centering
\includegraphics[width=\linewidth]{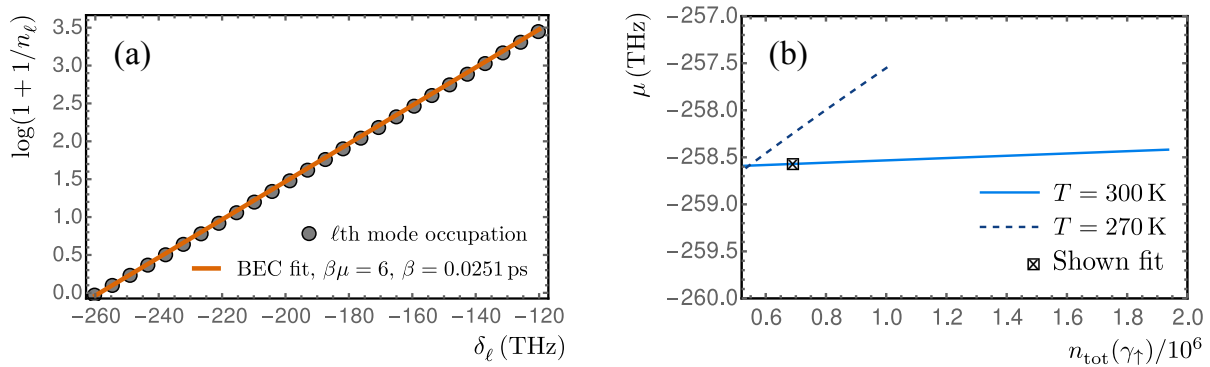}
\caption{ (a) Occupation $n_\ell^{}$ of every fourth mode for $\gamma_\uparrow = 0.011\gamma_\downarrow$, i.e., vertical triple-dashed line in figure~\ref{fig:occ-mu-plot}(a), shows linear behaviour in logarithmic representation indicating thermalisation. Fitted $\mu$ value shown as crossed square on the right. (b) Chemical potential $\mu$ as function of total photon number $n_{\rm tot}$  reveals non-zero slope for two different temperatures. We use $L=101$, other parameters are given in table~\ref{tab:exp_parameters}.}
\label{fig:occ-mu-plot-part2}
\end{figure}

The mode occupation $n_\ell^{}$ for a fixed $\gamma_\uparrow$ is shown in a logarithmic representation in figure~\ref{fig:occ-mu-plot-part2}(a), along the triple-dashed vertical line of figure \ref{fig:occ-mu-plot}(a). The linear behaviour indicates that the modes are distributed according to Bose-Einstein statistics
\begin{equation}
\label{eq:bec-distrib}
n_\ell^{} = \frac{1}{\exp[\beta(\delta_\ell - \mu)]-1} \, .
\end{equation}
Fitting a linear function to $\log \left(1 + 1 / n_\ell^{}\right)$ yields the inverse temperature $\beta$ and the chemical potential $\mu$ due to equation~(\ref{eq:bec-distrib}). For a non-interacting BEC condensate, i.e., $\mathcal{g}_\beta = 0$, the temperature obtained by fitting to a thermal cloud coincides with the spectral temperature of the dye and the chemical potential is locked to the value $\delta_1$ above the threshold \cite{Keeling_PRL-nonequilibrium_model_photon-cond}. However, here $\mathcal{g}_\beta$ is non-zero but small, which induces additional small corrections of the occupation. As a consequence, the effective thermalisation temperature of the thermal cloud differs generically from the spectral temperature of the dye. This behaviour depends on all system parameters, and especially a high pumping rate $\gamma_\uparrow$ can significantly change the temperature. For a choice of parameters outside of a certain region the non-equilibrium properties of the model are dominating the tendency to thermalise and the distribution is, in general, not thermal any more. This happens even in the case $\mathcal{g}_\beta = 0$ \cite{Keeling-Thermalization_photon_condensate}. Therefore, we restrict our further analysis only to the cases where a thermal cloud does exist. We have observed that a few of the lowest modes do not follow the linear dependency in figure~\ref{fig:occ-mu-plot-part2}(a), thus they are not considered in our fitting procedure. Additionally, we drop some of the highest modes as well, as they hold relatively small occupation and can, therefore, have large relative numerical error. The resulting value of $\mu$ from figure~\ref{fig:occ-mu-plot-part2}(a) is shown by a star symbol in figure~\ref{fig:occ-mu-plot-part2}(b). We repeat the procedure for different values of the pumping rate $\gamma_\uparrow$ and obtain a linear equation of state $\mu=\mu(n_{\rm tot})$, as presented in figure~\ref{fig:occ-mu-plot-part2}(b).

\subsection{Photon-photon interaction strength}

The slope $\partial \mu / \partial n_{\rm tot}$ of the equation of state $\mu=\mu(n_{\rm tot})$ in figure~\ref{fig:occ-mu-plot-part2}(b) is a consequence of the dye-mediated effective photon-photon interaction (see \ref{App:S-gbar-connection}), which is measured by the dimensionless interaction strength
\begin{equation}
\label{eq:gbar-slope-connection}
\tilde{g} = \frac{2\pi}{\hbar \Omega}\frac{\partial \mu}{\partial n_{\rm tot}}\,.
\end{equation}
Thus, we infer from figure~\ref{fig:occ-mu-plot-part2}(b) the resulting interaction strength $\tilde{g} = 5.2 \times 10^{-7}$, which agrees quite well with the range $\tilde{g}\sim 10^{-8}-10^{-7}$ given in \cite{StoofPRL-Interaction}. Note that, as mentioned in the introduction, the thermo-optical effect dominates the photon-photon interaction. Therefore, this value cannot be directly compared with the measured values \cite{Klaers_BEC_of_photons,NymanStrength,Oosten}, since we only model one of the respective contributions.

\begin{figure}[!t]
\centering
\includegraphics[width=0.77\linewidth]{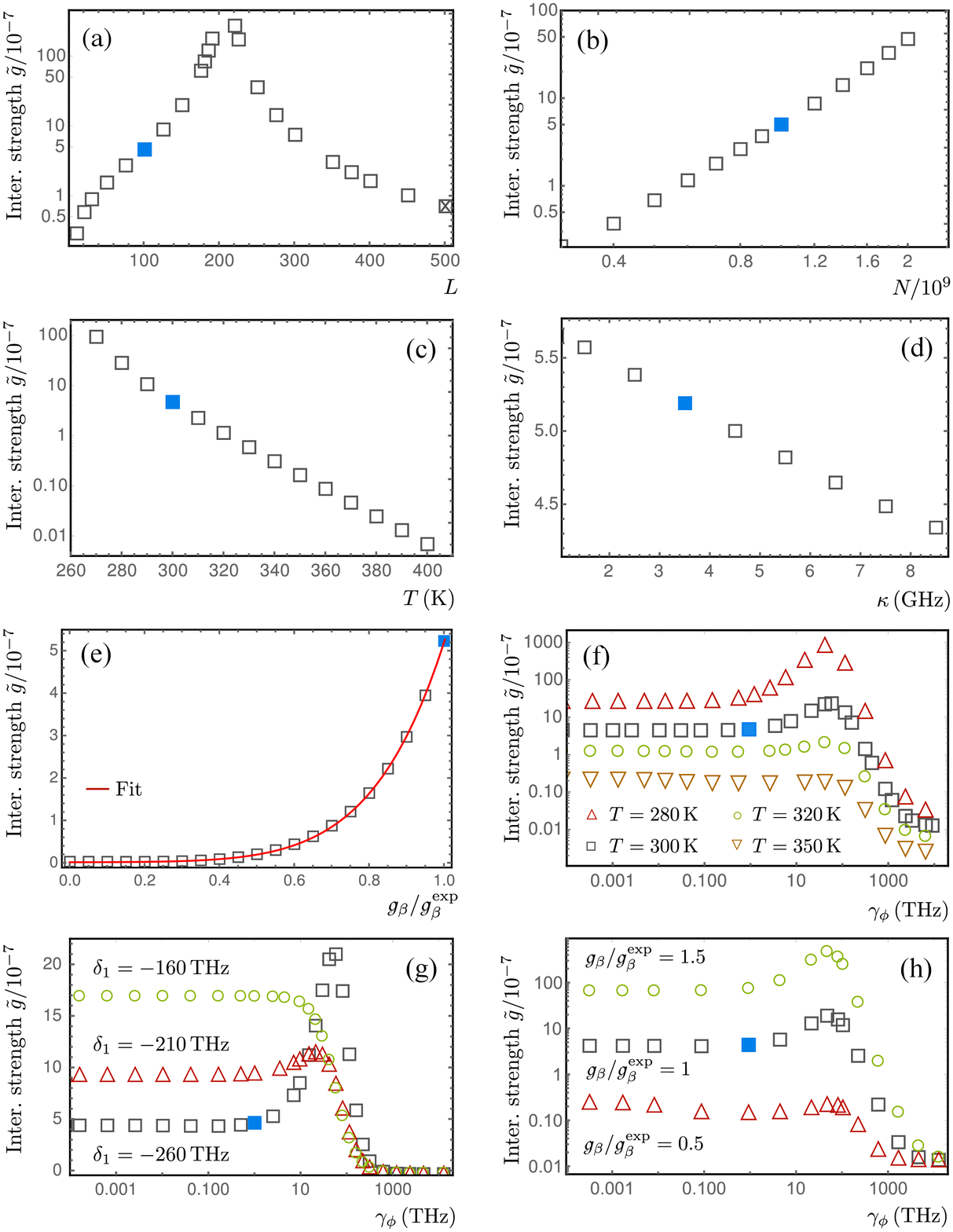}
\caption{
{Dependence of dimensionless effective photon-photon interaction strength $\tilde{g}$ on: (a) number of cavity levels $L$, (b) number of dye molecules $N$, (c) temperature $T$, (d) cavity decay rate $\kappa$ and (e) rescaled dressed dye-cavity coupling strength $\mathcal{g}_\beta/\mathcal{g}_\beta^{\rm exp}$, using a fixed $\gamma_\phi=0.1\,{\rm THz}$. Influence of dephasing rate for various: (f) temperatures, (g) cavity cutoffs $\delta_1$, (h) ratios $\mathcal{g}_\beta/\mathcal{g}_\beta^{\rm exp}$. From (a) we see that the number of cavity levels (i.e., the value of $\Omega$) in our model has a non-monotonous effect on the interaction strength $\tilde{g}$: the maximum is located around $L\approx 200$, where $L=501$ (crossed square) corresponds to the experimental regime. Blue filled squares in all panels show results in the case of experimentally chosen values (see table~\ref{tab:exp_parameters}), apart from $L=101$ which is used in (b)--(h). As we see, increase of $N$ or decrease of $T$ significantly increases $\tilde{g}$, whereas $\kappa$ and $\gamma_\phi$ affect its value only slightly in the vicinity of experimentally realistic values. Fit (red line) in (e) indicates the dependence of $\tilde{g}$ on $\mathcal{g}_\beta$ in the form of an even polynomial.}}\label{fig:gbar-depend}
\end{figure}

{Now we investigate how $\tilde{g}$ depends on the parameters of our model. First, we vary those that could be tuned in experimental setups: the number of cavity levels $L$ in the experimentally relevant range of $140\,{\rm THz}$ above $\delta_1$, the number of dye molecules $N$, the spectral temperature of the dye $T$ and the cavity decay rate $\kappa$. Next, it is instructive to think of $\mathcal{g}_\beta$ as being an independent parameter. In this way, by tuning $\mathcal{g}_\beta$ from zero to its experimental value $\mathcal{g}_\beta^{\rm exp}$, we are able to examine how the coherent terms of our model give rise to the effective photon-photon interaction strength. Finally, since the dephasing rate $\gamma_\phi$ is only approximately known in the experiments, we also investigate its influence on $\tilde{g}$ for various values of other parameters.} The corresponding results are presented in figure~\ref{fig:gbar-depend}. The  number of equations increases quadratically with the number of levels $L$. We have chosen $L=101$ due to computational constraints. Figure~\ref{fig:gbar-depend}(a) shows that the interaction strength $\tilde{g}$ depends non-monotonously on $L$. In case of up to 200 levels, the interaction $\tilde{g}$ increases nearly exponentially with the number of cavity levels. However, after reaching the maximum at $L\approx 200$ the dimensionless interaction strength $\tilde{g}$ starts to decrease. The value of $\tilde{g}$ in the case of the experimental number of levels $L=501$ is then comparable with the value for $L=101$ levels, which we used in our simulations.

According to figure~\ref{fig:gbar-depend}(b), a larger number of dye molecules $N$ increases $\tilde{g}$ dramatically, since more dye molecules mediating an effective coupling between the photons are present. In addition, much larger photon-photon interaction can also be achieved by lowering the temperature, see figure~\ref{fig:gbar-depend}(c). In contrast, figure~\ref{fig:gbar-depend}(d) reveals that increasing the decay rate $\kappa$ decreases only slightly the interaction strength $\tilde{g}$. An intuitive explanation of the last two results is as follows. Increasing either the temperature or the photon decay rate reduces the total number of photons in the system, thus there are less photons available to modify the dye medium and $\tilde{g}$ decreases correspondingly.

{In figure~\ref{fig:gbar-depend}(e) we investigate the dependency of $\tilde{g}$ on the coherent coupling $\mathcal{g}_\beta$, which we vary artificially from zero to the experimental value $\mathcal{g}_\beta^{\rm ext}$.  Quite expectedly, in the limit $\mathcal{g}_\beta\to 0$ the dimensionless interaction strength $\tilde{g}$ practically vanishes, the latter corresponding to the case of the Kirton-Keeling model \cite{Keeling_PRL-nonequilibrium_model_photon-cond,Keeling-Thermalization_photon_condensate}. The increase of $\tilde{g}$ is non-linear and in the considered range an even polynomial in $\mathcal{g}_\beta$ yields a good fit (red line). We observe that the quadratic term is almost negligible compared to quartic and higher-order terms, akin to the consideration of a photon-photon interaction corresponding to the box Feynman diagram analysed in the work \cite{StoofPRL-Interaction}. In our framework, such a dependence could easily be understood via a simple perturbative expansion of the expectation values $\avg{X}=\sum_{p=0}^\infty\avg{X}^{(p)}$,
where $\avg{X}^{(p)}\propto\mathcal{g}_\beta^p$, for an arbitrary system operator $X$ in the equations (\ref{eq:MainSys-1})--(\ref{eq:MainSys-5}), with the time derivatives set to zero. In such a way, higher perturbation orders can be systematically calculated from the lower ones. The zeroth order solutions for $\avg{n_m^{}}^{(0)}$ and $\avg{\sigma^z_1}^{(0)}$ are exactly those from the work of Kirton and Keeling \cite{Keeling_PRL-nonequilibrium_model_photon-cond,Keeling-Thermalization_photon_condensate}. It turns out that $\avg{n_m^{}}=\avg{n_m^{}}^{(0)}+\avg{n_m^{}}^{(2)}+\avg{n_m^{}}^{(4)}+\ldots$, so that we have an expansion $\tilde{g}=\tilde{g}^{(2)}+\tilde{g}^{(4)}+\ldots$ in even powers of $\mathcal{g}_\beta$.}

{Next, we analyse the effect of the dephasing on $\tilde{g}$ and the sensitivity of the obtained dependence on other system parameters. Figures \ref{fig:gbar-depend}(f)--\ref{fig:gbar-depend}(h) show that, quite generically, $\tilde{g}$ is affected insignificantly when the dephasing rate $\gamma_\phi$ is varied from zero to few ${\rm THz}$. Such a result could be attributed to the presence of a large detuning in (\ref{eq:MainSys-3}), of the order of $100\,{\rm THz}$, which anyhow strongly suppresses the coherent evolution on its own. Further increase of $\gamma_\phi$ may lead to the appearance of a resonance-like peak, followed by a polynomial decay for dephasing rates above several hundreds of THz. The latter is expected to happen when the dephasing rate becomes larger than $|\delta_1|$. The resonance-like peak becomes more pronounced as $\mathcal{g}_\beta$ is increased or the temperature decreased, as is seen in figures \ref{fig:gbar-depend}(f) and \ref{fig:gbar-depend}(h). This observation indicates that the peak is of coherent origin. On the other hand, the resonance-like peak can completely disappear when the cavity cutoff frequency is shifted towards the zero-phonon line, which is demonstrated in figure \ref{fig:gbar-depend}(g).}

\section{Conclusions}
\label{sec:con}

Here we presented a model which can interpolate between two different kinds of states of light in a micro-cavity, namely between a nearly non-interacting photon BEC and a laser-like state. Our model is based on a master equation approach, with an interplay between coherent and dissipative dynamics. The dominance of the former or the latter leads either to a coherent lasing state or to an equilibrated BEC state, respectively. We demonstrated that in the BEC case the lowest cavity energy level is macroscopically occupied and cavity modes of different energies are almost uncorrelated, whereas in the lasing case some cavity level becomes macroscopically occupied and strongly correlated with the others. Afterwards, we showed how to fix the parameters of our theory in an experimentally realistic regime. We emphasised that the coherent part of the master equation is then overwhelmed by the dissipative effects, but still large enough to lead to an additional effective photon-photon interaction. As a consequence, the chemical potential depends linearly on the total number of photons, as is expected from a perturbative solution of a Gross-Pitaevskii equation for the condensate wave function. This dependency allowed us to determine the dimensionless interaction strength $\tilde{g}$ to be of the order of $10^{-7}$ for experimentally realistic parameters.

We also investigated the dependency of $\tilde{g}$ on different model parameters, which can feasibly be tuned in the photon BEC experiments. Our numerics showed that increasing the number of dye molecules or decreasing the spectral dye temperature can significantly increase the $\tilde{g}$ value, whereas it is not much influenced by the cavity loss rate $\kappa$. However, this value cannot be directly connected to the current experimental values
\cite{Klaers_BEC_of_photons,NymanStrength,Oosten}. The reason is that in the experimental setups the dominating photon-photon interaction is of thermo-optical origin, whereas our theory has no spatial degrees of freedom and, thus, cannot capture such diffusive effects. Instead, in our case the effective photon-photon interaction could be compared with the dye-mediated photon-photon scattering. And indeed, our value is in the range of the expected estimate \cite{StoofPRL-Interaction}.

Another currently disputed feature of the photon condensate concerns its possible polarisation. Whereas no significant polarisation of the photon BEC was found in the original Bonn experiment \cite{Klaers_BEC_of_photons}, recent systematic measurements of the Stokes parameters in Utrecht \cite{NewOosten_2} indicate that the polarisation of the photon BEC correlates with the polarisation of the pump pulse. These new experimental results together with the recent theoretical investigation in the BEC case \cite{Keeling-Polarization} offer the prospect that the polarisation dependency could be investigated on the basis of an extension of our microscopic model during the whole crossover from the photon BEC to the laser-like phase.

\section*{Acknowledgements}

We acknowledge H. Haken, J. Keeling, P. Kirton, J. Kl\"ars, R. Nyman, D. van Oosten, G. Schaller, J. Schmitt, E. Stein, F. Vewinger and M. Weitz for inspiring discussions.
This work was supported in part by the Ministry of Education, Science and Technological Development of the Republic of Serbia under projects ON171017, ON171038, III45016 and BEC-L, by the German Academic and Exchange Service (DAAD) under project BEC-L, by the German Research Foundation (DFG) via the Collaborative Research Centers SFB 910, SFB/TR49 and SFB/TR185 and grant BR 1528/9-1 and by the European Commission through the project QUCHIP, grant No. 641039.

\begin{appendix}
\section{Interaction dependence of equation of state}
\label{App:S-gbar-connection}
Here we derive relation (\ref{eq:gbar-slope-connection}), which allows to determine the dimensionless photon-photon interaction strength $\tilde{g}$ from the slope
$\partial \mu / \partial n_{\rm tot}$ of the equation of state $\mu ( n_{\rm tot} )$.
To this end we follow \cite{Klaers_BEC_of_photons} and assume that the photon BEC is described by a condensate wave function $\Psi({\bf x})$, which obeys
a two-dimensional time-independent Gross-Pitaevskii equation:
\begin{equation}
\label{eq:gp-eq}
\left[-\frac{\hbar^2}{2m} \Delta + \frac{1}{2} m \Omega^2 {\bf x}^2  + g \left|\Psi({\bf x})\right|^2\right] \Psi({\bf x}) = \mu \Psi({\bf x})\,.
\end{equation}
Here, $g$ denotes the photon-photon interaction strength, $m$ stands for the photon mass, $\Omega$ is the trapping frequency and $\mu$ represents the chemical potential. As the photon-photon interaction strength
$g$ is supposed to be small, we solve equation~(\ref{eq:gp-eq}) perturbatively. At first we neglect the interaction $g$, so equation~(\ref{eq:gp-eq}) can be solved exactly.
The ground-state wave function $\Psi^{(0)}({\bf x})$, which is normalised to the total number of photons $n_{\rm tot}$, reads
\begin{equation}
\label{eq:gp-eq-gs-wavefunction-ohne-interaction}
\Psi^{(0)}({\bf x}) = \sqrt{\frac{n_{\rm tot}}{\pi l^2}} \exp\left(- \frac{{\bf x}^2}{ 2 l^2}\right),
\end{equation}
with the oscillator length $l=\sqrt{\hbar/ m \Omega}$ and the chemical potential $\mu^{(0)} = \hbar \Omega$ coincides with the zero-point energy of the two-dimensional harmonic oscillator.
For a non-vanishing interaction strength $g$ we assume a perturbative correction in first order for both the condensate wave function and the chemical potential:
\begin{eqnarray}
\Psi({\bf x}) &=& \Psi^{(0)}({\bf x}) + \Psi^{(1)}({\bf x}) + \ldots \, ,\\
\mu &=& \mu^{(0)} + \mu^{(1)}+ \ldots \, . 
\end{eqnarray}
With this ansatz the Gross-Pitaevskii equation (\ref{eq:gp-eq}) reduces to 
\begin{equation}
  \label{pert}
\left[-\frac{\hbar^2}{2m} \Delta + \frac{1}{2}m \Omega^2 {\bf x}^2 - \mu^{(0)} \right] \Psi^{(1)}({\bf x}) = \mu^{(1)} \Psi^{(0)}({\bf x}) - g \Psi^{(0)}({\bf x})^3\,, 
\end{equation}
which determines both interaction corrections $\Psi^{(1)}({\bf x})$ and $\mu^{(1)}$. In our context it is sufficient to calculate the latter one, which follows from the Fredholm alternative \cite{Fredholm}. To this end we multiply equation~(\ref{pert}) with $\Psi^{(0)}({\bf x})$ and integrate over ${\bf x}$, so we get due to (\ref{eq:gp-eq-gs-wavefunction-ohne-interaction}) with 
the dimensionless interaction parameter \cite{Dalibard}
\begin{equation}
\label{eq:interaction_dimensionlos_def}
\tilde{g} = \frac{g m}{\hbar^2}
\end{equation}
the equation of state
\begin{equation}
\label{eq:chem_pot_gl_dimlos}
\mu = \hbar \Omega + \frac{\tilde{g} \hbar \Omega}{ 2\pi} n_{\rm tot} + \ldots \,.
\end{equation} 
Thus, for small interactions $\tilde{g}$ the chemical potential $\mu$ changes linearly with the photon number $n_{\rm tot}$. The slope $\partial \mu / \partial n_{\rm tot}$ of the equation of state $\mu ( n_{\rm tot} )$
depends then via $\partial \mu / \partial n_{\rm tot} = \tilde{g} \hbar \Omega / (2\pi)$ linearly on the dimensionless interaction strength $\tilde{g}$, which leads to relation (\ref{eq:gbar-slope-connection}).
\end{appendix}

\end{document}